\documentclass[galaxies,article,accept,moreauthors,pdftex,10pt,a4paper]{mdpi}

\renewcommand{\epsilon}{\varepsilon}

\usepackage{epsfig,amsmath,wrapfig}

\newcommand{\dF}{{^{^*}\!\!F}}
\newcommand{\bP}{{\bf P}}
\newcommand{\bF}{{\bf F}}
\newcommand{\bU}{{\bf U}}

\newcommand{\sB}{{\mathcal{B}}}

\newcommand{\del}{{\partial}}

\firstpage{1}
\articlenumber{x}
\doinum{10.3390/------}
\pubvolume{xx}
\pubyear{2016}
\copyrightyear{2016}
\externaleditor{Academic Editor: name}
\history{Received: date; Accepted: date; Published: date}

 \theoremstyle{mdpi}
 \newcounter{thm}
 \newcounter{ex}
 \newcounter{re}
\Title{Black hole accretion in gamma ray bursts}

\Author{Agnieszka Janiuk $^{1}$*, Michal Bejger $^{2}$, Petra Sukova $^{1}$, and Szymon Charzynski $^{1,3}$}
\AuthorNames{Agnieszka Janiuk, Michal Bejger, Petra Sukova and Szymon Charzynski}

\address{%
$^{1}$ \quad Center for Theoretical Physics, Polish Academy of Sciences, Al. Lotnikow 32/46, 02-668 Warsaw, Poland; agnes@cft.edu.pl\\
$^{2}$ \quad Nicolaus Copernicus Astronomical Center, Polish Academy of Sciences, ul. Bartycka 18, 00-716 Warsaw, Poland; bejger@camk.edu.pl \\
$^{2}$ \quad Chair of Mathematical Methods in Physics, University of Warsaw, ul. Pasteura 5, 02-093 Warsaw, Poland
}

\corres{Correspondence: agnieszka.janiuk@gmail.com}

\abstract{
We study the structure and evolution of the hyperaccreting disks and outflows in the gamma ray bursts central engines. The torus around a stellar mass black hole is composed of free nucleons, Helium, electron-positron pairs, and is cooled by neutrino emission. Accretion of matter powers the relativistic jets, responsible for the gamma ray prompt emission. The significant number density of neutrons in the disk and outflowing material will cause subsequent formation of heavier nuclei. We study the process of nucleosynthesis and its possible observational consequences.
We also apply our scenario to the recent observation of the gravitational wave signal, detected on September 14th, 2015 by the two Advanced LIGO detectors, and related to an inspiral and merger of a binary black hole system. A gamma ray burst that could possibly be related with the GW150914 event was observed by the Fermi satellite. It had a duration of about 1 second and appeared about 0.4 seconds after the gravitational-wave signal. We propose that a collapsing massive star and a black hole in a close binary could lead to the event. The gamma ray burst was powered by a weak neutrino flux produced in the star remnant's matter. Low spin and kick velocity of the merged black hole are reproduced in our simulations. Coincident gravitational-wave emission originates from the merger of the collapsed core and the companion black hole.
}


\begin{document}


\section{Introduction}

Gamma-ray bursts (GRBs) are energetic, transient events observed
in the sky at high energies. Their known cosmological origin
requires a physical process that causes them to be a cosmic explosion of
great power. Proposed mechanisms involve the creation of a black hole (BH) in a
cataclysmic event. This may either result from the collapse of a massive
rotating star, or via the merger of compact objects' binary, e.g. neutron stars (NSs) or a BH and a neutron star. The so-called `central engine' of this
process is composed of a hot and dense accretion disk with a hyper-Eddington
accretion rate (up to a few $M_\odot s^{-1}$) near a spinning BH.

In the hyperaccreting disks that are present around the BHs in the central engine, the densities and temperatures are so high, that the equation of state (EOS) can no longer be assumed to be of an ideal gas.
The pressure and chemical balance is required by the nuclear reactions that
take part between the free protons, neutrons, and electron-positron pairs abundant in the plasma. The charged particles must satisfy the total
 neutrality condition.
In the $\beta$ reactions, neutrinos are produced in three flavors and are subject to absorption and scattering processes. The partially trapped neutrinos
contribute to the pressure in the plasma, together with nucleons, electron-positron pairs, radiation, and synthesized heavier particles (Helium).

Numerical computations of the structure and evolution of the
accretion flows in the gamma ray bursts engine were historically first carried out
with the use of steady-state, and later with the time-dependent models.
These models were axially and vertically averaged, and used the classical $\alpha$-parameter prescription for the viscosity
(Popham et al. 1999; Di Matteo et al. 2002; Kohri et al. 2002, 2005; Chen \& Beloborodov 2007; Reynoso et al. 2006; Janiuk et al. 2004; 2007; 2010).
More recently, accretion flows are described by fully relativistic
MHD computations (e.g., Barkov \& Komissarov 2011; Janiuk et al. 2013).
This method has also been applied recently to describe
the central engine of a putative gamma ray burst which could be associated with
the event GW150914 (Janiuk et al. 2017). Here we report on this work,
and in addition, we also present new results on the nucleosynthesis of
heavy elements in the accretion flow in this engine.

\section{Accretion flow in the GRB engine}

\subsection{Equation of state}

In the EOS, contribution to the pressure comes from free nuclei and $e^{+}-e^{-}$ pairs, helium, radiation and trapped
neutrinos:
\[P = P_{\rm nucl}+P_{\rm He}+P_{\rm rad}+P_{\nu}, \]
where $P_{\rm nucl}$ includes free neutrons, protons,
and the electron-positrons:
\[P_{\rm nucl}=P_{\rm e-}+P_{\rm e+}+P_{\rm n}+P_{\rm p}\]
with
\[P_{\rm i} = {2 \sqrt{2}\over 3\pi^{2}}
{(m_{i}c^{2})^{4} \over (\hbar c)^{3}}\beta_{i}^{5/2}
\left[F_{3/2}(\eta_{\rm i},\beta_{\rm i})+{1\over 2} \beta_{\rm i}F_{5/2}(\eta_{\rm i},\beta_{\rm i})\right].\]
Here, $F_{\rm k}$ are the Fermi-Dirac integrals of the order $k$, and
$\eta_{\rm e}$, $\eta_{\rm p}$ and $\eta_{\rm n}$ are the reduced chemical
potentials, $\eta_i = \mu_i/kT$
is the degeneracy
parameter, $\mu_i$ denoting the standard chemical
potential. Reduced chemical potential of positrons is
$\eta_{\rm e+}=-\eta_{\rm e}-2/\beta_{\rm e}$.
Relativity parameters are defined as $\beta_{\rm i}=kT/m_{\rm i}c^{2}$.
The term $P_{\rm rad}$, describing the pressure of radiation,
scales with the temperature as $T^{4}$, and is added here together with the
pressure of neutrinos, $P_{\nu}$, which is computed from the two-stream approximation (see details in Janiuk et al. 2007). The disk is fully opaque to photons, however the radiation pressure is still by several orders of magnitude smaller than the pressure of neutrinos.
This EOS is computed numerically by solving the balance of nuclear reactions (Yuan 2005; Janiuk et al. 2007; Janiuk \& Yuan 2010; see also Lattimer \& Swesty 1991; Setiawan et al. 2004).

\subsection{Neutrino cooling of the accretion flow in GRBs}

In the hot and dense torus, with temperature of ${\sim}10^{11}$ K
and density $> 10^{10}$ g cm$^{-3}$, neutrinos are efficiently produced.
The main reactions that lead to their emission are the electron/positron capture on nucleons,
as well as the neutron decay. Their nuclear equilibrium is described by the following
reactions:
\begin{eqnarray}
\label{eq:urca}
p + e^{-} \to n + \nu_{\rm e} \nonumber \\
p + \bar\nu_{\rm e} \to n + e^{+}  \nonumber \\
p + e^{-} + \bar\nu_{e} \to n \nonumber \\
n + e^{+} \to p + \bar\nu_{\rm e} \nonumber \\
n \to p + e^{-} + \bar\nu_{\rm e} \nonumber \\
n + \nu_{\rm e} \to p + e^{-},  \nonumber
\end{eqnarray}
and the forward and backward reaction rates are equal.
The reaction rates are given by the appropriate integrals
(Reddy, Prakash \& Lattimer 1998; see also Appendix A in Janiuk et al. 2007).

Other neutrino emission processes are: electron-positron pair annihilation, bremsstrahlung and plasmon decay:

\begin{equation}
e^{-}+e^{+}\to \nu_{\rm i}+\bar\nu_{\rm i} \nonumber \\
\label{eq:annihil}
\end{equation}
\begin{equation}
n+n \to n+n+\nu_{\rm i}+\bar\nu_{\rm i} \nonumber \\
\label{eq:brems}
\end{equation}
\begin{equation}
\tilde \gamma \to \nu_{\rm e}+\bar\nu_{\rm e} . \nonumber \\
\label{eq:plasmon}
\end{equation}
We calculate their rates numerically, with proper integrals over the distribution function of relativistic, partially degenerate species.

The neutrino cooling rate is finally given by the two-stream approximation, and includes
the scattering and absorptive optical depths for neutrinos of all three
flavors (Di Matteo et al. 2002):
\[Q^{-}_{\nu} = { {7 \over 8} \sigma T^{4} \over
{3 \over 4}} \sum_{i=e,\mu,\tau} { 1 \over {\tau_{\rm a, \nu_{i}} + \tau_{\rm s} \over 2}
+ {1 \over \sqrt 3} +
{1 \over 3\tau_{\rm a, \nu_{i}}}} \times {1 \over H}\; ~[{\rm erg ~s^{-1} ~cm^{-3}]} .\]
The neutrino cooling is self-consistenlty computed from the balance of the nuclear reactions, listed in Section 2.2, whose rates govern the values of
the absorptive optical depths for electron (all reactions) and muon or tau neutrinos (pair annihilation and bremsstrahlung reactions). Also, the neutrino scattering optical depth is computed, using the values of proton and neutron number densities, and Cabbibo angle of $\sin^{2}\theta_{\rm C}=0.23$ (see Yuan 2005, Janiuk et al. 2007). The emissivity is averaged over the disk height, instead of the integration over the neutrinosphere, whose shape is rather complex.

\subsection{GR MHD scheme}

Our simulations of the accretion flow dynamics in the GRB central engine
were performed with a 2D code {\it HARM}
(High Accuracy Relativistic Magnetohydrodynamics;  Gammie et al. 2003).
Our version of the code was extended to include the microphysics, described in the previous Section.
The code provides a solver for the continuity and energy-momentum conservation equations:
\[ (\rho u^\mu)_{;\mu} = 0 \]
\[ {T^\mu}_{\nu;\mu} = 0 \]
\[ P = K\rho^\gamma = (\gamma-1) u \]
where:
\[ T^{\mu\nu} = T^{\mu\nu}_{gaz} + T^{\mu\nu}_{EM} \]
\[ T^{\mu\nu}_{gaz} = \rho h u^\mu u^\nu + pg^{\mu\nu} =(\rho + u + p) u^\mu u^\nu + pg^{\mu\nu} \]
\[ T^{\mu\nu}_{EM} = b^2 u^\mu u^\nu + \frac{1}{2}b^2 g^{\mu\nu} - b^\mu b^\nu ; ~~ b^\mu = u_{\nu}\dF^{\mu\nu} \]
where $u^\mu$ is the four-velocity of gas, $u$ is the internal energy density,
$b^\mu = \frac{1}{2} \epsilon^{\mu\nu\rho\sigma}u_\nu F_{\rho\sigma}$ is the magnetic four-vector,
and $F$ is the electromagnetic stress tensor. The metric tensor is denoted by $g^{\mu\nu}$.
Within the force-free approximation, we have $E_{\nu}=u_{\mu}F^{\mu\nu}=0$.

{\it HARM-2D} uses a conservative numerical scheme to obtain solutions of equations
of the following type:
\[\del_t \bU(\bP) = -\del_i \bF^i(\bP) + \mathbf{S}(\bP),\]
where $\bU$ denotes a vector of ``conserved'' variables (momentum, energy density,
number density, taken in the coordinate frame), $\bP$ is a vector of
'primitive' variables (rest mass density, internal energy),
 $\bF^i$ are the fluxes, and $\mathbf{S}$ is a vector of source terms.
In contrast to non-relativistic MHD, where $\bP \rightarrow \bU$ and  $\bU
\rightarrow \bP$ have a closed-form solution, in relativistic MHD,
$\bU(\bP)$ is a complicated, nonlinear relation. Inversion
$\bP(\bU)$ is calculated therefore numerically, at every time step.
The inverse transformation requires a solution to 5 non-linear equations,
done here by means of a multi-dimensional Newton-Raphson routine
(see, e.g., Noble et al. 2006 for details).

The procedure is simple, if the pressure-density relation is adiabatic.
However, for a general EOS, one must also compute $dp/dw$, $dp/dv^{2}$
and $p(W, v^{2},D)$, where $w=p+u+\rho$ denotes the enthalpy, $W=\gamma^2 w$, $D=\sqrt{-g}\rho u^t$, and $v^{2}=g^{\mu\nu}u^{\mu}u^{\nu}$.
In our scheme, we have tabulated $(p,u)(\rho,T)$ values and then interpolate over this table.
The Jacobian $\partial \bU /\partial \bP$ is computed numerically. The conserved variables are then evolved in time.

{\it HARM-2D} has been MPI-parallelized for the hydro-evolution, and also implemented with the shared memory hyperthreading for the EOS-table interpolation.
Our 2D simulations typically run up to t=2000-3000 M (within a couple weeks of computations on the local cluster, or a few days on Cray XC40 with 1k nodes).
The computations are performed in the polar coordinate system, with
resolution of 256 x 256 points in the $r$ and $\theta$ directions.

\subsection{Example simulation results}

Parameters used in our computations are: BH mass, with a fiducial value of $M=10 M_{\odot}$, BH spin $a=0.6-0.98$, and accretion disk mass $M_{d}\approx 1 M_{\odot}$, which gives the scaling for the density in physical units,
needed for the EOS computations.
To describe the rotationally-supported torus around a spinning BH
we use the initial condition given by the equilibrium solution, first developed analytically by Fishbone \& Moncrief (1976) and Abramowicz et al. (1978).
We adopt the polar coordinate system,
$r-\theta$, and use the Kerr-Schild coordinates to avoid the coordinate singularity on the BH horizon. Initially, a poloidal configuration of the field is assumed, with the vector potential $A_{\phi}= (\rho/\rho_{max})$, and initial normalization of
$P_{gas}/P_{mag}=50$. Magnetic turbulences develop within the dense matter torus
and the field is advected with the infalling gas under the BH horizon.
For a rapidly-spinning BH, the open magnetic field lines form along the rotation axis, and the magnetically driven jets can be produced. Energy extracted from the rotating BH through the Blandford-Znajek process (Blandford \& Znajek 1977)
gives then a substantial contribution to the jet luminosity. In addition, annihilation of neutrino-antineutrino pairs produced in the torus is yet another source of power to the jets. In Figure 1, we show the result of an exemplary simulation, which shows a snapshot taken from evolved run, with the distribution of magnetic field lines, magnetic and gas pressure ratio, and neutrino emissivity, in the region close to a black hole.

\begin{figure}[H]
\centering
\includegraphics[width=6cm,angle=270]{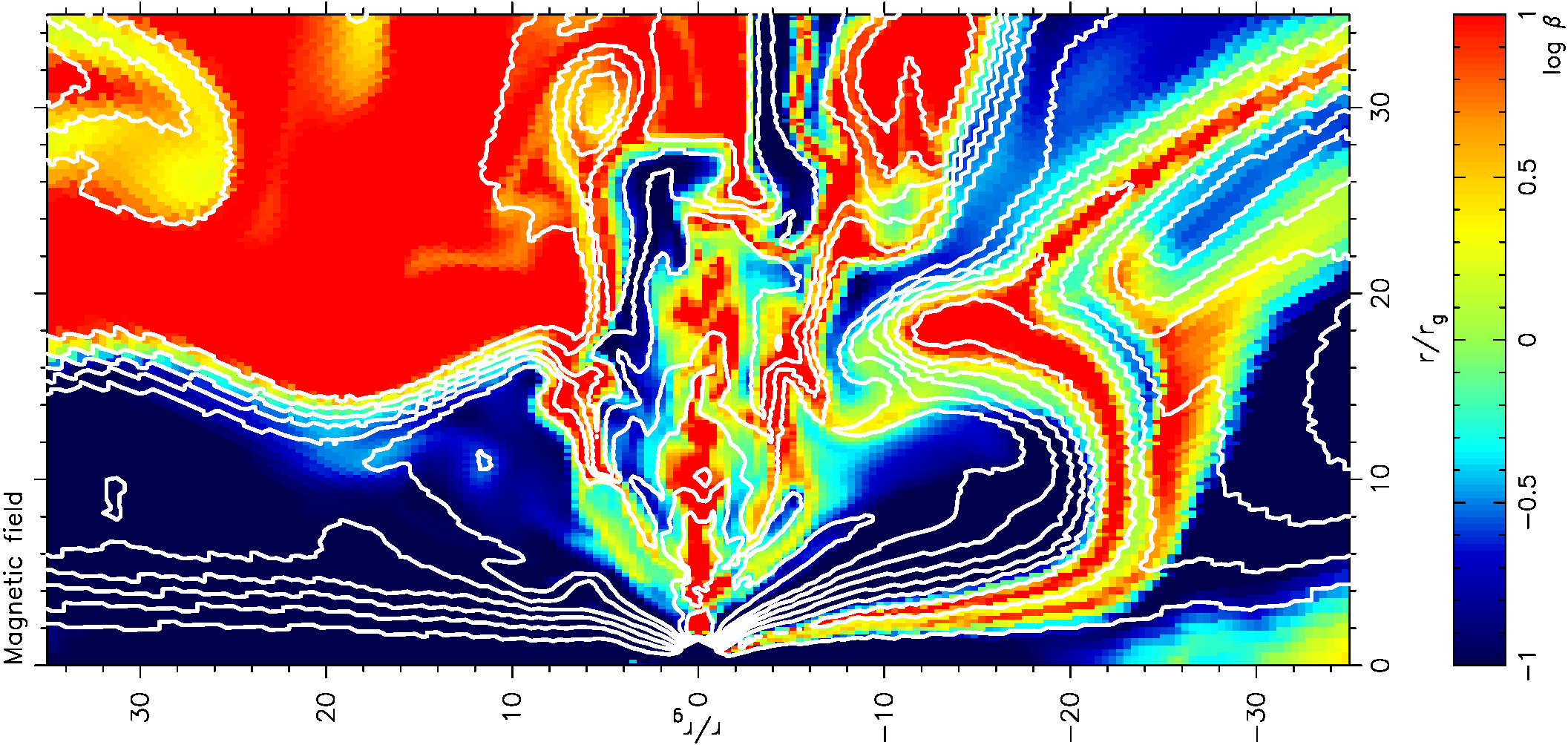}
\includegraphics[width=6cm,angle=270]{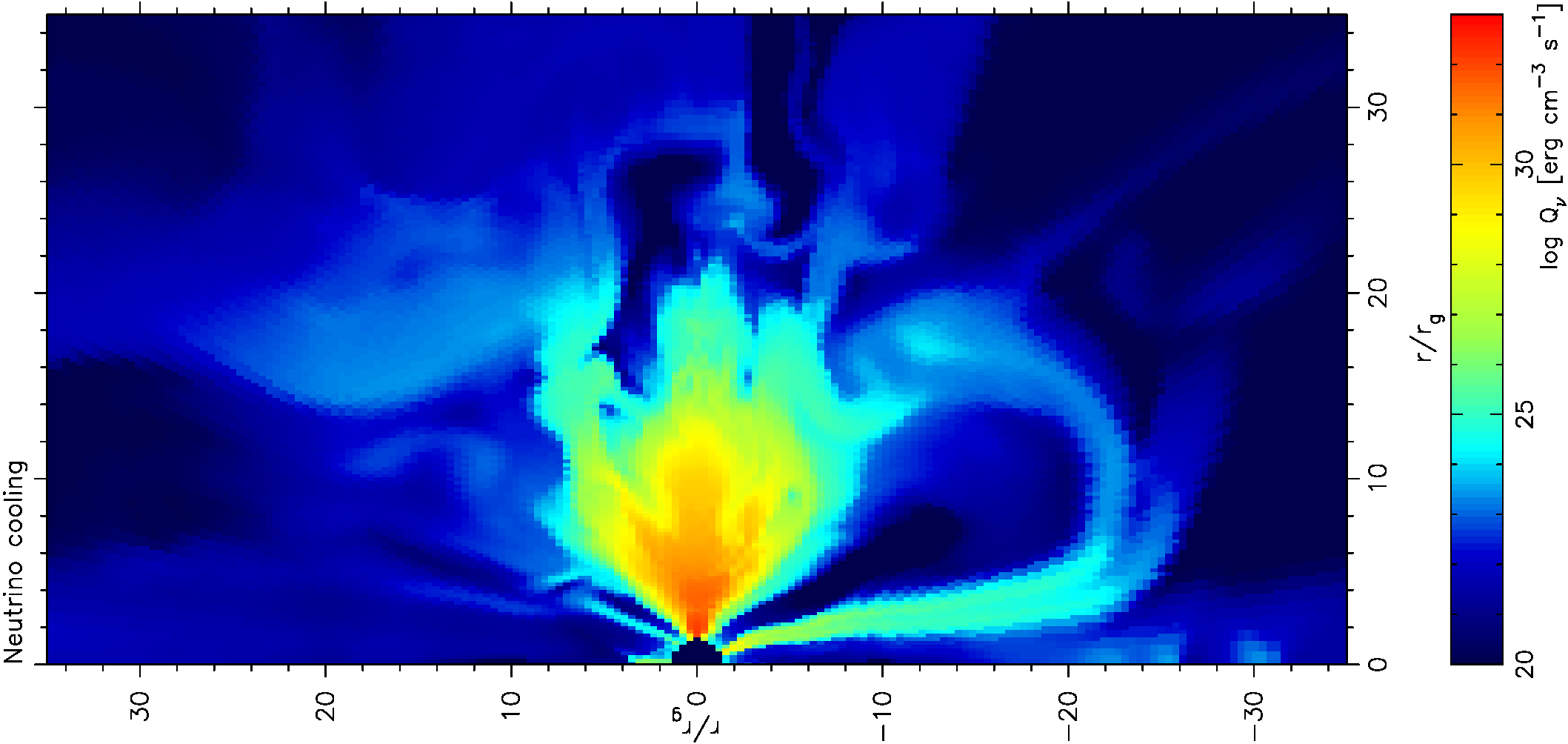}
\caption{Magnetisation $\beta=P_{gas}/P_{mag}$ and magnetic field lines topology
(left), and neutrino emissivity distribution (right) in the evolved
accretion flow. Snapshots taken at time $t=2000 M$. Black hole spin is $a=0.98$.}
\end{figure}

\section{Nucleosynthesis of heavy elements in the GRB engine}

In the astrophysical plasma, thermonuclear fusion occurs
due to the capture and release of particles ($n$, $p$, $\alpha$, $\gamma$).
 Reaction sequence produces further isotopes.
The nuclear reactions may proceed with one (decays, electron-positron capture, photodissociacion), two (encounters), or three (triple alpha reactions) nuclei.
Therefore, the change of abundance of the $i$-th isotope is in general given by:
\begin{equation}
\dot Y_{\rm i} = \sum_{j} N_{j}^{i} \lambda_{j}Y_{j} +
\sum_{j,k} N_{j,k}^{i} \rho N_{\rm A} Y_{j}Y_{k} +
\sum_{j,k,l} N_{j,k,l}^{i} \rho^{2} N^{2}_{\rm A}Y_{j}Y_{k}Y_{l} \nonumber
\end{equation}
Abundances of the isotopes are calculated under the assumption of nucleon number and charge conservation for a given density, temperature and electron fraction ($T \le 1 MeV$). Integrated cross-sections depending on temperature $kT$ are determined with the Maxwell-Boltzmann or Planck statistics, and the background screening and degeneracy of nucleons must be taken into account.
The set of resulting non-linear differential equations is solved using the Euler method (Wallerstein et al. 1997).

In our computations, we used the thermonuclear reaction network code, \textit{http://webnucleo.org}, and we computed the nuclear statistical equilibria established for the fusion reactions, for which the data were taken from the \textit{JINA reaclib} online database (Hix \& Meyer 2006).
This network is appropriate for temperatures below $1$ MeV, which is the case at the outer radii of accretion disks in GRB engines.
The mass fraction of the isotopes is solved for converged profiles of density,
temperature and electron fraction in the disk.

\begin{figure}[H]
\centering
\includegraphics[width=6cm]{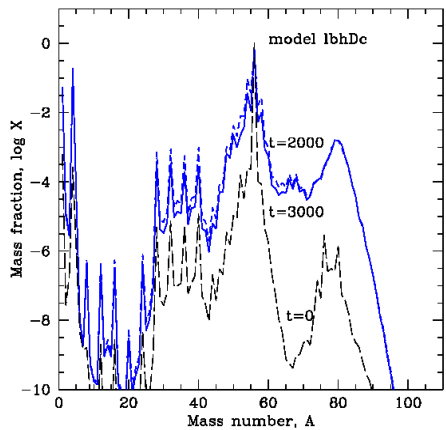}
\caption{
Volume integrated abundance distribution of elements
synthesized in the
accretion disk, based on the 2D simulation. Parameters: disk to BH mass ratio of
$M_{\rm d}/M_{\rm BH} = 0.25$, BH spin $a=0.9$. Abundances were calculated
from the density, temperature, and electron fraction profiles, as evolved to the time $t=2000 M$ (solid line) and $t=3000 M$ (dashed line), while the long-dashed line shows the distribution at time $t=0$. }
\end{figure}

We find that the free nuclei are present mostly below 100-200 gravitational radii, and the plasma there is rather neutron rich.  Then, most abundant heavy elements formed in the accretion tori in GRBs are Nickel, Iron and Cobalt, as well as  Argonium, Titatnium, Cuprum, Zinc, Silicon,
Sulphur, Clorium, Manganium, Titatnium, and Vanadium.

As a consequence of the heavy element formation, the enhanced emission in the lower energy bands, i.e., ultraviolet or optical, due to the decay of species, may accompany the GRB (the 'macronova'; e.g., Li \& Paczynski 1998).
Also, the radio flares, occurring months to years after the GRB can be observed (Piran et al. 2012).
Furthermore, certain isotopes decay should be detectable via emission lines.
For the \textit{NuSTAR} satellite, sensitive in the $5-80$ keV range,
it's in principle possible to detect the photons from Ti, Co, Mn, Cu, Zn, Ga, Cr unstable isotopes decays.
Additionally, the EPIC detector onboard the XMM-Newton,
could be able to see lines below 15 keV, e.g., $^{45}Ti$, $^{57}Mn$,
or $^{57}Co$ (Janiuk 2014).
In Figure 3, we plot an example result of the nucleosynthesis computations,
showing the volume integrated abundance distribution of elements, in the function of their mass number. The input was taken from the simulation of the GRB central engine, as designed to model the putative GRB that could possibly be associated with GW 150914 (see next Section).

\section{Gravitational wave source GW150914}

Data from Fermi Gamma-ray Burst Monitor (GBM) satellite suggested that the recently registered gravitational wave event GW150914 (Abbott et al. 2016), a coalescing binary BH system, is potentially related to a GRB. The electromagnetic radiation in high energy (above 50 keV) originated from a weak transient source and lasted for about 1 second (Connaughton et al. 2016). Its localization is broadly consistent with the sky location of GW150914.

We speculate here on the possible scenario for the formation of a GRB accompanied by the gravitational wave (GW) event. Even though the
presence of the GRB was recently questioned by other instruments measurements (e.g., Greiner et al. 2016), we envisage the possibility of a GRB coincident with the GW event from the BH merger to be worth exploring, in the context of the future observations.

Our model invokes coalescence of a collapsing star's nucleus that forms a BH, with its companion BH in a binary system (Janiuk et al. 2013, 2017).
We find that the recoil velocity acquired by the final BH through the GW emission allows it to take only a small fraction of matter from the host star, provided specific configuration of the binary spin vectors with the orbital angular momentum. The GRB is produced on the cost of accretion of this remnant matter onto the final BH. The moderate spin of this BH accounts for the GRB jet to be powered by a weak neutrino emission rather than the Blandford-Znajek mechanism, and hence agrees with the low power observed in the Fermi GRB signal.

The semi-analytical treatment of the BH core collapse and the star's
envelope spin-up (see Janiuk et al. 2008, for details) is followed by the general-relativistic simulations of the binary BH merger. Then, the GRB central engine evolution is carried numerically, to find the energy output from the magnetized, neutrino-cooled torus that powers the electromagnetic burst.

\subsection{Merger of binary BHs} \label{GW_merger}

The GW150914 event was interpreted to be a merger of two BHs of the
masses of $36^{+5}_{-4}\ M_\odot$ and $29^{+4}_{-4}$ (Abbott et al. 2016). The final BH parameters were estimated
to be of $62^{+4}_{-4}\ M_\odot$ and $0.67^{+0.05}_{-0.07}$ for its mass and
spin. Probabilities that the angles between spins and the normal to the orbital
plane are between $45^\circ$ and $135^\circ$ are about $0.8$ for each component
BH. Spin magnitudes are constrained to be smaller than $0.7$ and $0.8$ at 90\%
probability. Assumption of a strict co-alignment of spins with the
orbital angular momentum results in an upper limit of $0.2$ and $0.3$ for the
spins. Distance to the source was that of
$410^{+160}_{-180}$ Mpc, corresponding to a redshift of about $z=0.09$ (assuming
standard cosmology).

\begin{figure}[H]
\centering
\includegraphics[width=7cm]{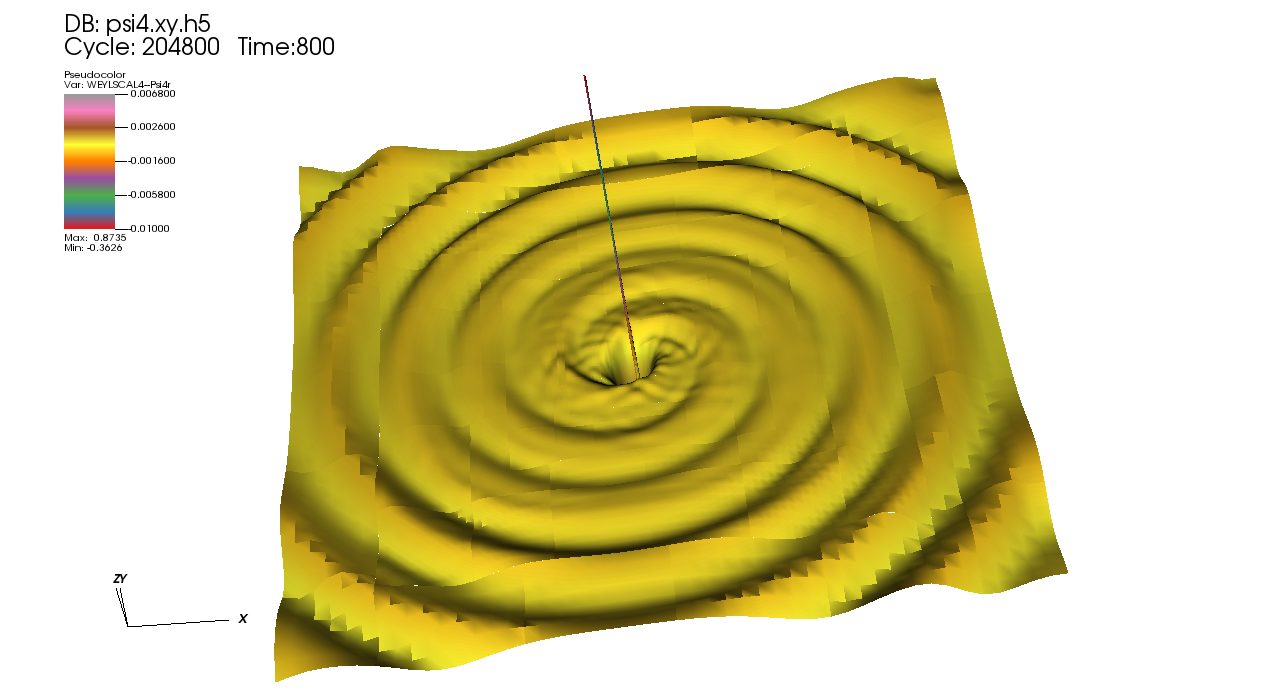}
\includegraphics[width=5cm]{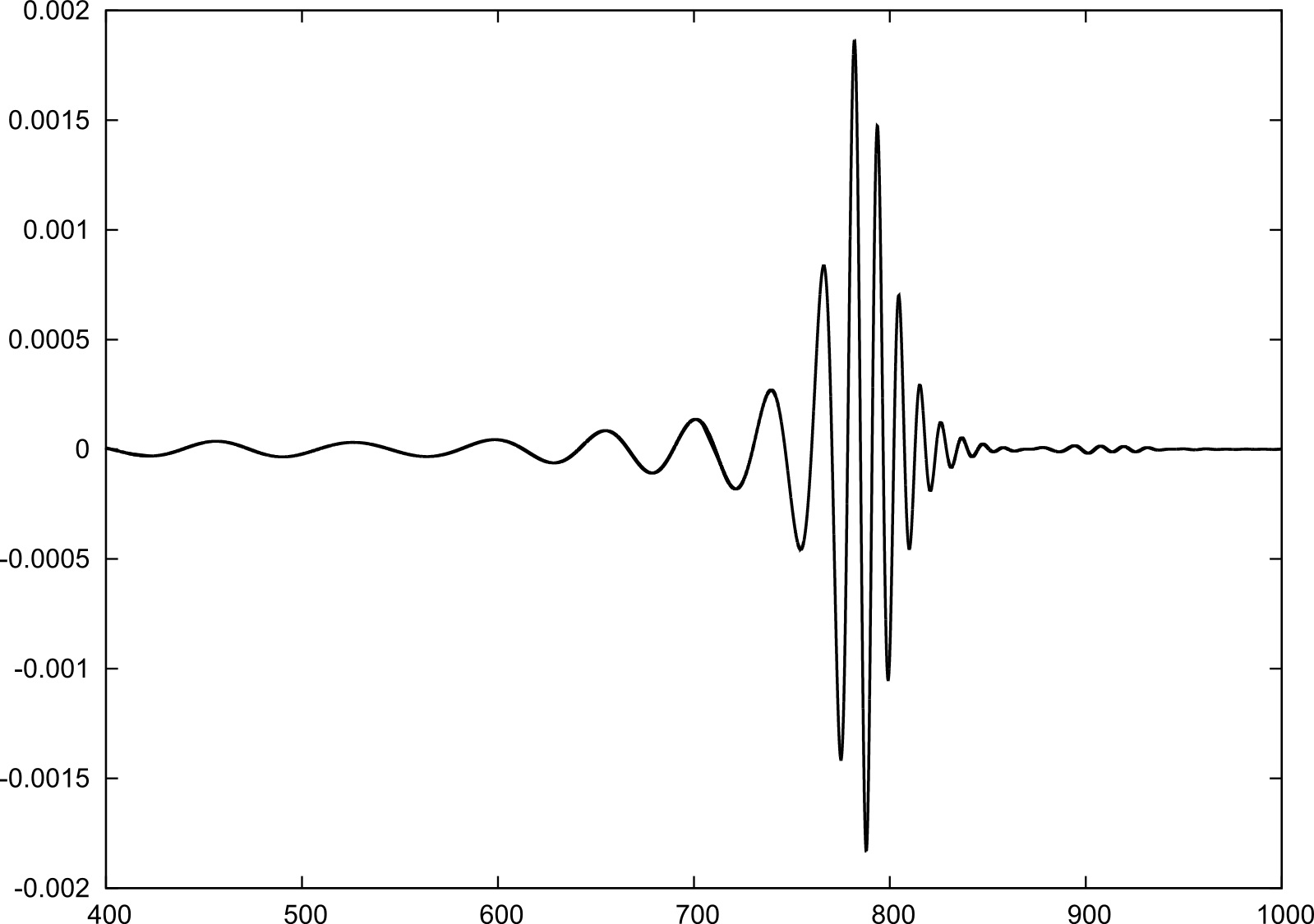}
\caption{Visualization of the spacetime curvature just after the black hole merger, described by the Weyl scalar  $\psi_4$, which shows the outgoing radiation for asymptotically flat spacetime (left), and an exemplary GW signal that can be detected in the instrument, shown by the time dependence of the real part of the $l=2$, $m=2$ components of $\psi_4$ (right).
}
\end{figure}
We made several runs for BH mass and range of spins, constrained from the Advanced LIGO data. In Figure 3, we plot the example results from the run with BH mass ratio of $0.82$, and aligned spins of $0.2$ and $0.3$. The waveform is plotted using the Weyl scalar $\psi_4$ \footnotemark. The final BH spin obtained in this simulation was $0.68$.

\footnotetext{Curvature of spacetime is described by curvature tensor - the Riemman tensor (20 independent components), which can be decomposed into sum of Ricci tensor (10 independent components) and Weyl tensor (10 independent components).
For vacuum solutions of Einstein equations Ricci tensor vanishes, so curvature is described by Weyl tensor only.
In the Newman-Penrose formalism components of Weyl tensor are encoded as five complex Weyl scalars $\Psi_{0}$, ... , $\Psi_{4}$. These are different components of curvature tensor, which have different physical interpretation.
$\Psi_{4}$ describes the outgoing radiation for an asymptotically flat spacetime.}

 These runs were performed with the Einstein Toolkit computational package
(Loeffler et al 2012).
The numerical methods used here are based on finite difference computations on a
gridded mesh and follow the inspiral, merger and ringdown phases of the binary black hole evolution, using the technique of the adaptive mesh
refinement. The Cartesian grid covers the volume of $48\times48\times48 M$. In AMR we use 7 levels of refinement, and the
resolution ranges from $\Delta
x=2.0M$ for the coarsest grid to $\Delta x=0.03125M$ for the finest grid.
From the simulation runs, we constrained not only the spin of the merger black hole (it is then used to be a parameter in the GRB engine model), but also the
velocity of the gravitational recoil. This turned out to be on the order of $\sim700$ km/s. Following Kocsis et al. (2012), we argue that it is likely that accretion of the accumulated matter onto the final BH is triggered while it moves towards the circumbinary disk after the merger.

\subsection{Weak GRB powered by neutrino emission}

The Fermi GRB coincident with the GW150914 event had a duration of about 1 sec and appeared about 0.4
seconds after the GW signal. Within the limit of uncertainty of the Advanced LIGO and Fermi detectors' capabilities it could also be associated spatially.
The GRB fluence in the range 1 keV-10
MeV, is of $2.8 \times 10^{-7}$ erg cm$^{-2}$.
The implied source luminosity in gamma rays equals to $1.8^{+1.5}_{-1.0} \times 10^{49}$ erg/s.

We modeled the GRB engine, constrained by the post-merger conditions. Using the code {\it HARM-2D}, with the numerical EOS and neutrino cooling implemented into MHD evolution, we estimated the neutrino and Blandford-Znajek luminosities available to power the GRB jet in this source.
In particular, the mass of the accreting torus, created from the circumbinary matter, was tuned to produce adequately low neutrino luminosity for a weak GRB.
Example parameters for the BH mass of $M=62 M_{\odot}$, the BH spin of $a=0.7$, and disk mass $M_{d}=15 M_{\odot}$ are consistent wit the weak GRB luminosity.
Assuming low efficiencies of conversion between
neutrino annihilation, and weak conversion between the
jet kinetic and radiative power, this scenario is able to meet quantitatively
the limits put by the Fermi data.
The disk mass is here just an order of magnitude estimate. We first checked, that a much larger mass, on the order of the mass of the merged black hole, is too large to be consistent with the observed limitations for luminosity. On the other hand, the supernova explosion which might have left a 30 Solar mass black hole, should have originated from at least 80 Solar mass star on zero-age main sequence, in a low-metallicity environment (Abbott et al. 2016b, Spera et al. 2015). Even if a significant amount of the star's mass was ejected during the evolution and explosion, some remnant of this order is plausible.

\begin{figure}[H]
\centering
\includegraphics[width=5cm]{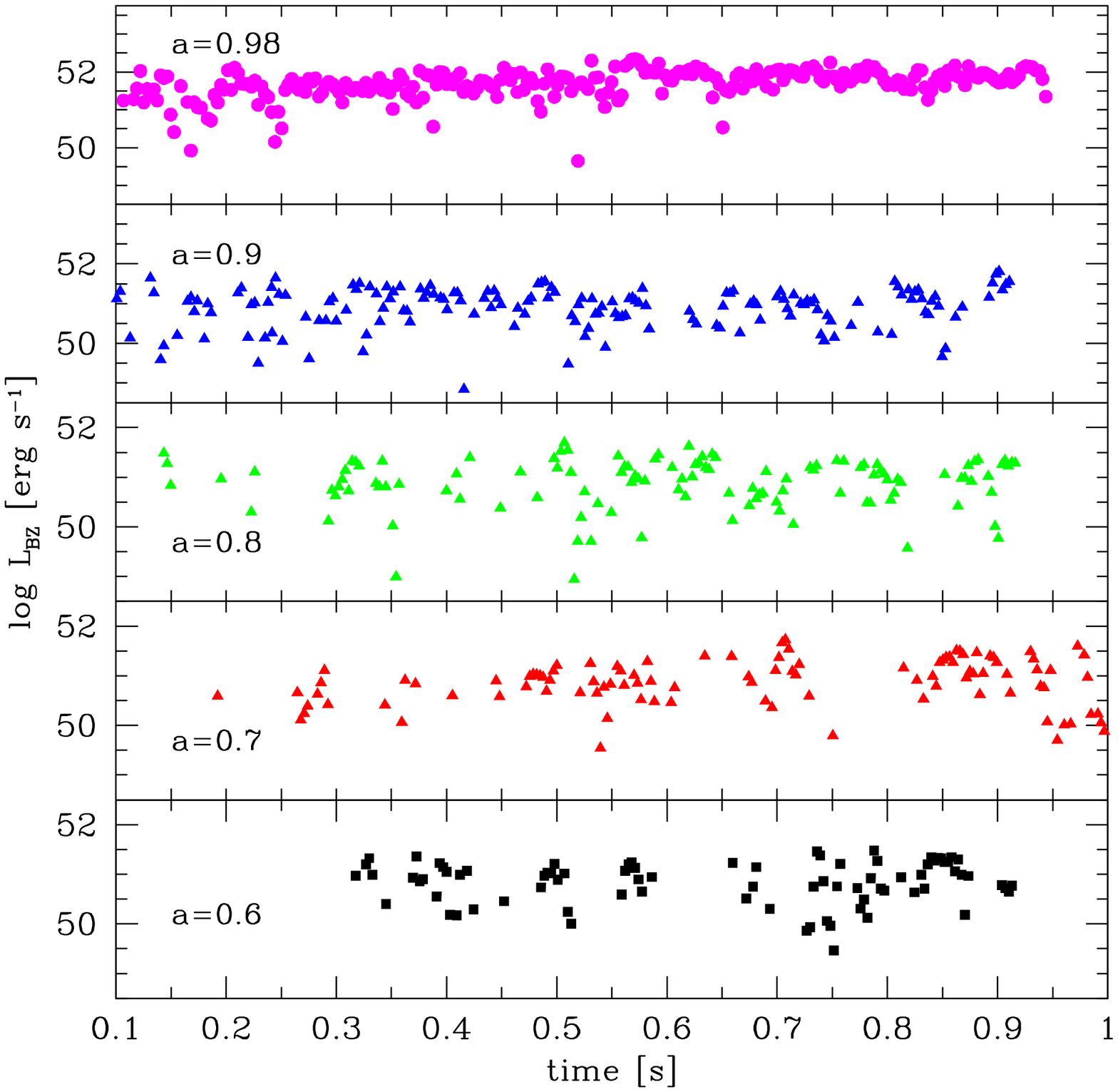}
\includegraphics[width=5cm]{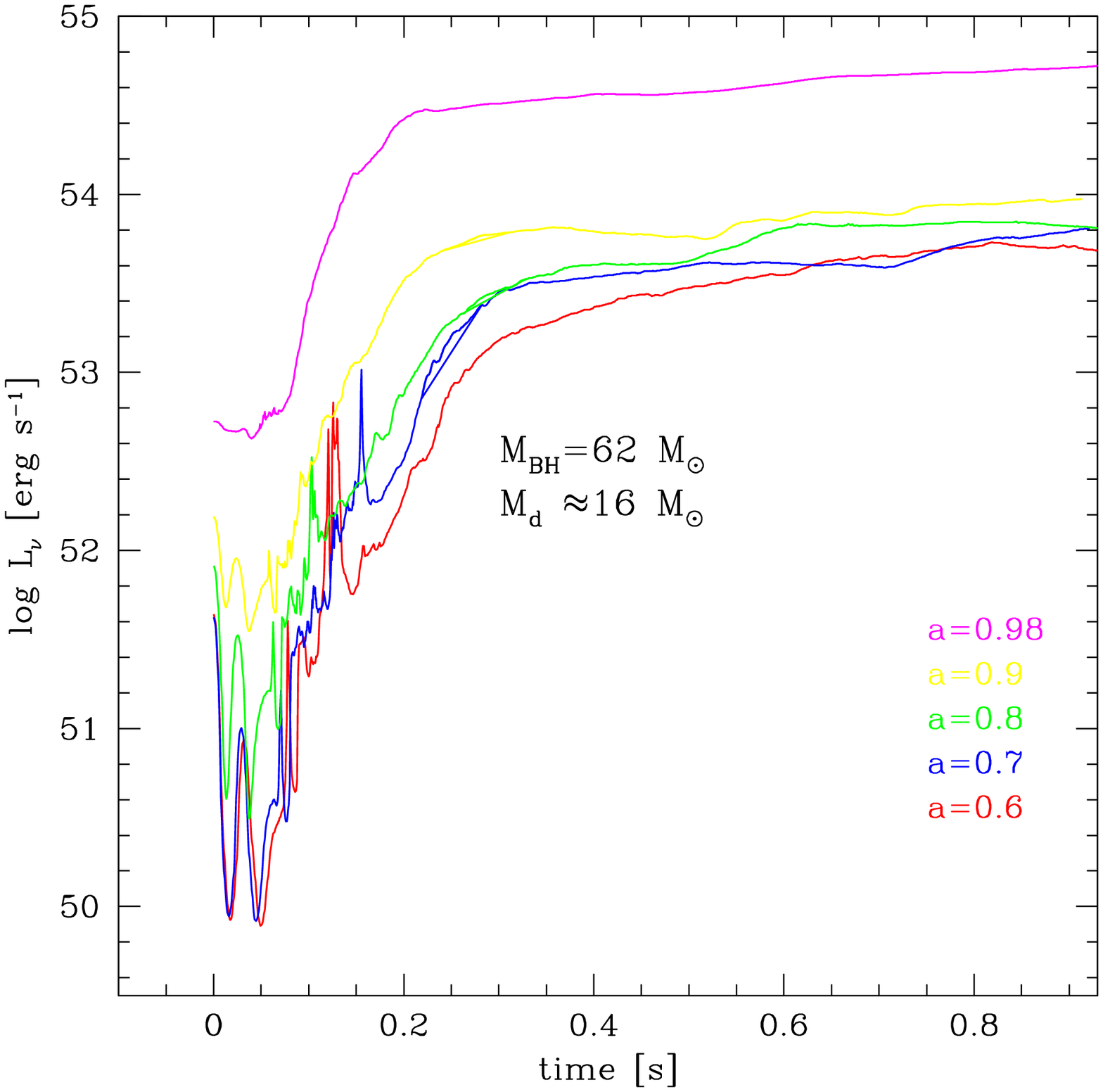}
\caption{Power available in the GRB engine through the neutrino emission (right), and through the  magnetic field advected under the BH horizon (left).
}
\end{figure}

The luminosity $L_{\nu}$, emitted in neutrinos is simply equal to their volume integrated emissivity.
To compute the energy flux through the horizon of the BH and the Blandford-Znajek luminosity, $L_{\rm BZ}$, we need
the electromagnetic part of the stress tensor,
\begin{equation}
T^{\mu\nu}_{\rm EM} =
b^2 u^\mu u^\nu + {b^2\over{2}} g^{\mu\nu} - b^\mu b^\nu,
\end{equation}
where the four-vector $b^\mu$ with $b^t \equiv
g_{i\mu} B^i u^\mu$ and $b^i \equiv (B^i + u^i b^t)/u^t$.
We then evaluate the radial energy flux as the power of the Blandford-Znajek process:
\begin{equation}
L_{\rm BZ} \equiv \dot{E} = 2\pi\int_0^{\pi} \, d\theta \,
{\sqrt{-g}}{F_E}
\end{equation}
where $F_E \equiv -{{T^{r}_t}}$.  This can be subdivided into a matter ${F}^{(MA)}_E$
and electromagnetic ${F}^{(EM)}_E$
part, although in the
force-free limit the matter part vanishes (McKinney \& Gammie 2004).

In Figure 4, left panel, we plot the values of $L_{\rm BZ}$ as it varies with time for various simulations, depending on the black hole spin. In the right panel, we show the corresponding evolution of the neutrino luminosity.
The maximum of the neutrino luminosity
obtained in our simulations is reached about 0.4 seconds after an equilibrium
torus, prescribed by our initial conditions, had formed. This may tentatively
give the lower limit for the
timescale when the jet appears after the BH merger. The Blandford-Znajek luminosity is on the order of $10^{50}-10^{51}$ ergs/s, and only occasionally non-zero for low spins (the jet is powered by magnetic field only episodically).
We find therefore, that the GRB luminosity inferred from our simulations
can be reconciled with the observational
upper limits, for moderate spins of the final BH ($a=0.6-0.8$).

The connection of our computation to the particular event GW150914 and the corresponding Fermi GRB is arguable because of several facts. At first, the flare produced by our simulation is longer than the reported very short duration of the GRB (1s). However, the observed burst was close to the sensitivity limit of the instrument, so it is likely that it could observe just the very top of the flare and left the rest of the burst undetected. Moreover, if our line of sight is offset with respect to the jet axis and we can see only the edge of the jet, the detailed shape of the flare would be affected by the behaviour of the propagating jet. The time delay between the burst and gravitational signal can be significantly longer, if there is a massive star remnant, through which the jet has to propagate. The state of the remnant matter after the second BH collapse is very uncertain and needs to be studied in details in future works. However, we can speculate that the collapse and subsequent orbiting of the two BHs is violent enough to destroy the star, expel part of the matter away and form a dead circumbinary disc of the remnant matter, in which the accretion has stopped, around the BH binary. Then after the merger the new born black hole can receive a substantial kick towards the circumbinary disc (see our BH merger simulations in Section \ref{GW_merger}), can meet the matter in fraction of second and produce the burst. In that case, there is only little material left along the axis, through which the jet has to propagate, and so the time delay can be quite short, on the order of a second. However, this speculative scenario needs both more theoretical modeling and mainly more future observations of coincident gravitational and electromagnetic signals, which are well above the instrumental limits.

\section{Conclusions}

\begin{itemize}
\item We have carried out numerical GR MHD simulations of the magnetized torus accreting onto BH in the GRB central engine. The EOS is no longer assumed to be of an ideal gas, but accounts for the proper microphysics and neutrino cooling of the GRB engine.
\item The neutrino emission and absorption processes account for an additional pressure component in the plasma, and the species are partially degenerate. The resulting density and temperature, as well as the electron fraction, are then taken into account when determining the nuclear equilibrium conditions.
\item We determine the abundances of heavy elements (above Helium, up to the mass number of $\sim 100$), which are synthesized in the accretion disk in GRB engine, and in the winds that are magnetically launched from its surface. The observational signatures of radioactive decay of some of the unstable isotopes can be, for instance, the emission lines seen in the X-rays (in principle, in the \textit{NuSTAR} range),
and also in the faint emission in lower energy band continuum (i.e. 'kilonova' or 'macronova' scenario).
\item We compute and compare the efficiencies of the GRB jet powered through the neutrino annihilation, and the Blandfrod-Znajek mechanism. We find that the constraints of a moderate BH spin and a small disk mass to the BH mass ratio, are in tentative agreement with the Advanced LIGO and Fermi data, if the latter was really coincident with the GW event.
\item A definite answer and a test for our scenario will be possible if further searches for the GW events will provide more data, also for their electromagnetic counterparts, in both gamma rays and lower energies.
\end{itemize}
We claim, that even though the observations are still subject to debate, the possible GRB-BBH merger coincidence is worth to investigate. The timescales
obtained in our scenario are possibly more plausible for a lonng gamma ray burst scenario than for a short one (see also Perna et al. 2016; Loeb 2016), unless
the Fermi burst itself was much longer, but below the detction limit. On the other hand, the model proposed by Zhang (2016) can satisfy the duration and delay timsecales, but the intepretaion of a charged black hole and its observational signatures also needs further investigations.

\vspace{6pt}

\acknowledgments{This work was supported in part by the grants no. DEC-2012/05/E/ST9/03914 and
UMO-2014/14/M/ST9/00707 from the Polish National Science Center.
We also acknowledge support from the Interdisciplinary Center for Mathematical Modeling of the Warsaw University, through the computational grant G53-5.}

\bibliographystyle{mdpi}

\renewcommand\bibname{References}


\end{document}